\documentclass[aps,prl,twocolumn,superscriptaddress,showpacs]{revtex4-1}

\usepackage{amssymb}
\usepackage{amsmath}
\usepackage{graphicx}
\usepackage{natbib}
\usepackage{nicefrac}
\usepackage{multirow}
\usepackage{epstopdf}
\usepackage{float}
\usepackage{physics}

\begin{document}

%\title{Probing Valley Splitting in Si Quantum Dots via Circuit Quantum Electrodynamics}
%\title{Microwave Spectroscopy of Valley States in Si Quantum Dots}
\title{High Resolution Valley Spectroscopy of Si Quantum Dots}

\author{X.~Mi}
\affiliation{Department of Physics, Princeton University, Princeton, New Jersey 08544, USA}
\author{Csaba~G.~P\'{e}terfalvi}
\affiliation{Department of Physics, University of Konstanz, D-78464 Konstanz, Germany}
\author{Guido~Burkard}
\affiliation{Department of Physics, University of Konstanz, D-78464 Konstanz, Germany}
\author{J.~R.~Petta}
\affiliation{Department of Physics, Princeton University, Princeton, New Jersey 08544, USA}

\pacs{03.67.Lx, 73.21.La, 42.50.Pq, 85.35.Gv}
% 03.67.Lx Quantum computation
% 73.21.La Quantum dots
% 42.50.Pq Cavity quantum electrodynamics; micromasers
% 85.35.Gv Single electron devices

\begin{abstract}
We study an accumulation mode Si/SiGe double quantum dot (DQD) containing a single electron that is dipole coupled to microwave photons in a superconducting cavity. Measurements of the cavity transmission reveal dispersive features due to the DQD valley states in Si. The occupation of the valley states can be increased by raising temperature or applying a finite source-drain bias across the DQD, resulting in an increased signal. Using cavity input-output theory and a four-level model of the DQD, it is possible to efficiently extract valley splittings and the inter- and intra-valley tunnel couplings.
\end{abstract}

\maketitle

Spin states of electrons in gate-defined semiconductor quantum dots (QDs) are among the leading qubit candidates in efforts to build a solid state quantum processor \cite{Loss_DiVincenzo_PRA}. Since spin qubits are often indirectly manipulated through electrical means to achieve fast control \cite{Petta_Science,HRL_Nature2012}, a precise knowledge of the other quantum degrees of freedom governing QD electrons is of critical importance for improving the quantum gate fidelities. In III-V semiconductors such as GaAs, these relevant quantum degrees of freedom include charge and orbital states, which are reproducibly defined by QD lithographic dimensions and may be spectroscopically probed through a variety of techniques, such as photon assisted tunneling and pulsed-gate spectroscopy \cite{Kouwenhoven_Review_2002,Elzerman2004}.

In silicon, which supports long spin lifetimes and is therefore highly suited for spin qubit implementations \cite{Lyon_NatMat_2012,VeldhorstM.2014,Dave_Ninedot_PhysRevApp}, electrons possess an additional quantum degree of freedom. The conduction band of bulk Si has six degenerate minima (termed valleys) \cite{RevModPhys.54.437}. In Si/SiGe heterostructures, the four in-plane valleys are raised in energy compared to the two out-of-plane valleys through the strain in the Si quantum well \cite{Schaffler_SiGe_Review,RevModPhys.85.961}. The relatively small energy splitting between the two low-lying valley states has been observed to contribute to spin relaxation \cite{Yang_NatComm_2013,Kawakami.2014}, but may potentially also be harnessed to make charge-noise-insensitive qubits  \cite{DasSarma_ValleyQubit_2012}. This valley splitting has been found to vary substantially within the range of 35 -- 270 $\mu$eV in Si/SiGe QD devices \cite{Borselli_ValleyAPL_2011,Dave_DQD_APL}, posing an urgent challenge to the reproducibility and scalability of spin qubits based on Si/SiGe heterostructures. A first step toward controlling valley splitting is the development of an experimental method to accurately and efficiently determine its value.

In this Letter, we demonstrate a cavity-based method of measuring valley splitting in Si/SiGe DQDs using a hybrid circuit quantum electrodynamics (cQED) device architecture \cite{Mi_APL_2016,Mi_Science_2016}. Charge transitions involving excited valley states generate observable ``fingerprints'' in the stability diagram of a cavity-coupled Si/SiGe DQD, as predicted by a recent theory \cite{PhysRevB.94.195305}. The occupation of the valley states can be increased by raising the device temperature or by applying a finite source-drain bias, $V_{\rm SD}$, across the DQD. Such a cavity-based valley detection scheme is highly efficient since it eliminates the need for a magnetic field. Our approach yields information on the valley splitting, intra- and inter-valley tunnel couplings, and is therefore an attractive alternative to conventional magnetospectroscopy and photon assisted tunneling \cite{Ke_PRL,PhysRevB.94.195305}.

\begin{figure}[t]
	\centering
	\includegraphics[width=\columnwidth]{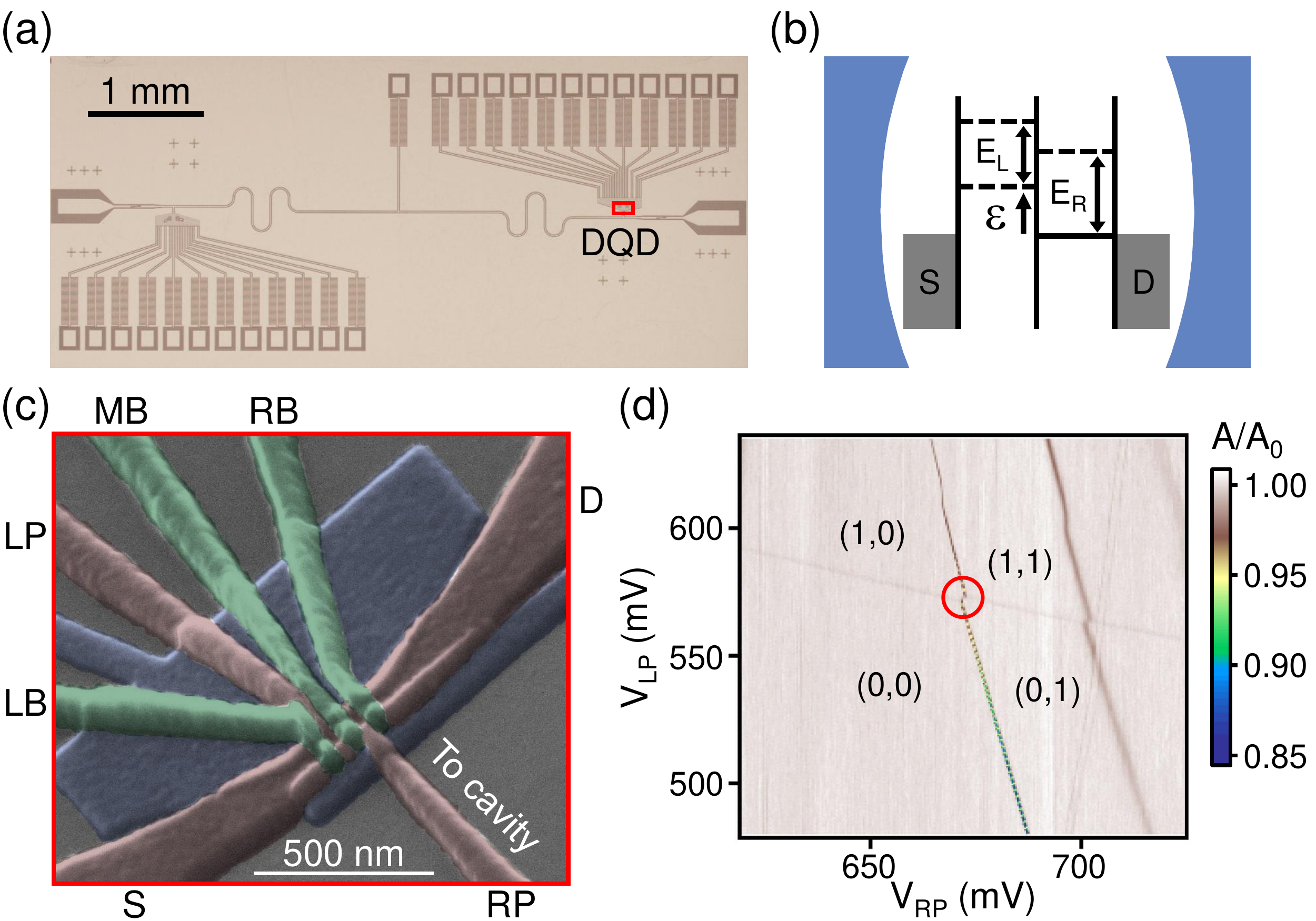}
	\caption{(a) Optical image of the device. (b) Schematic representation of the experiment and the DQD energy levels. $E_\text{L}$ and $E_\text{R}$ denote the valley splittings in the left and right dots, and $\epsilon$ is the interdot level detuning. (c) Tilted angle false-colored scanning electron microscope image of the DQD [red outline in (a)]. (d) DQD charge stability diagram acquired by measuring the cavity transmission amplitude $A/A_0$ as a function of the plunger gate voltages $V_\text{LP}$ and $V_\text{RP}$, with fixed interdot barrier gate voltage $V_\text{MB} = 150$ mV.}
	\label{fig:1}
\end{figure}

The hybrid Si/SiGe cQED device is shown in Fig.~\ref{fig:1}(a). The cavity is a half-wavelength ($\lambda/2$) Nb transmission line resonator with a center frequency $f_\text{c} = 7.796$ GHz, loaded quality factor $Q_\text{c} = 2480$, and photon loss rate $\kappa / 2\pi = 3.1$ MHz. A DQD is defined in a Si quantum well near a voltage anti-node of the cavity [Fig.~\ref{fig:1}(c)]. Three overlapping layers of Al gates are patterned on top of an undoped Si/SiGe heterostructure to achieve tight electronic confinement \cite{Dave_DQD_APL,Mi_APL_2016}. The Si/SiGe heterostructure consists of a 4 nm thick Si cap, a 50 nm thick Si$_{0.7}$Ge$_{0.3}$ spacer layer, a 8 nm thick Si quantum well, and a 225 nm thick Si$_{0.7}$Ge$_{0.3}$ layer grown on top of a Si$_{1-x}$Ge$_{x}$ relaxed buffer substrate.

Figure~\ref{fig:1}(b) shows a schematic representation of the equilibrium configuration of the DQD energy levels. In contrast with previous III/V semiconductor DQD-cQED devices, where two charge states interact with the cavity photons, a total of four charge states are involved in the charge-cavity interaction in this work due to the presence of valley states in Si \cite{Walraff_2012_PRL,Petersson_Nature_2012}. Electric dipole coupling between DQD electrons and cavity photons is maximized by connecting gate RP to the cavity center pin \cite{Walraff_2012_PRL,Mi_APL_2016}. 

Readout of the DQD charge states is performed at the base temperature of a dilution refrigerator ($T_{\rm lat}$ = 10 mK) by driving the cavity with a coherent microwave tone at frequency $f = f_\text{c}$ and power $P_\text{in} \approx -128$ dBm (the average intra-cavity photon number $n_c \approx 1$). The cavity output field is amplified and demodulated to yield the normalized transmission amplitude $A/A_0$ and phase $\Delta \phi$ response \cite{Petersson_Nature_2012,Mi_APL_2016,Supplement}. Figure~\ref{fig:1}(d) shows $A/A_0$ as a function of plunger gate voltages $V_\text{LP}$ and $V_\text{RP}$, revealing a few-electron DQD charge stability diagram \cite{Walraff_2012_PRL,Mi_APL_2016}. Charge stability islands are labeled with $(N_L,N_R)$, with $N_L$ and $N_R$ being the total number of electrons in the left dot and the right dot.

\begin{figure}[t]
	\centering
	\includegraphics[width=\columnwidth]{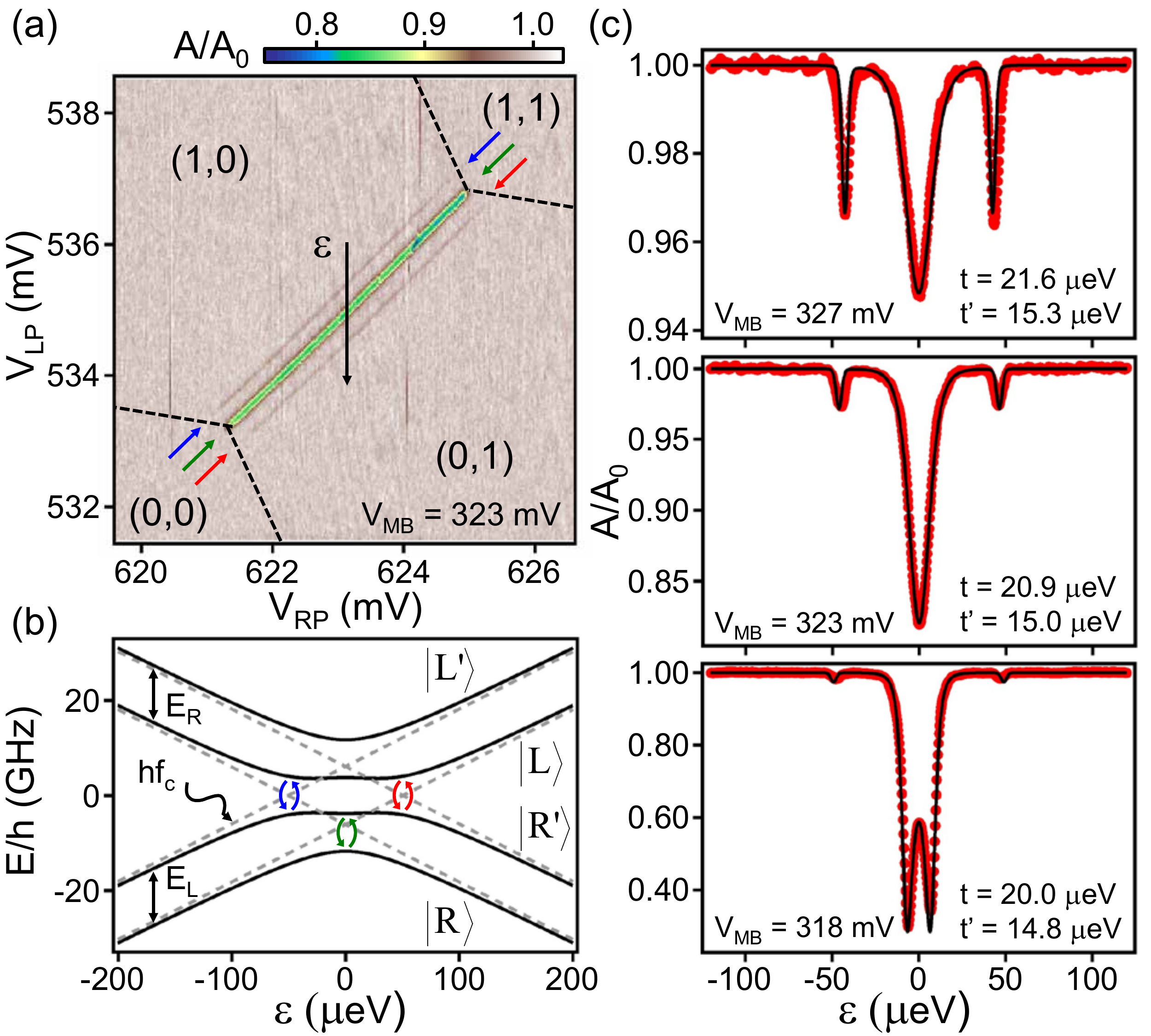}
	\caption{(Color Online). (a) Cavity transmission amplitude $A/A_0$ plotted as a function of $V_\text{LP}$ and $V_\text{RP}$ near the $(1,0) \leftrightarrow (0,1)$ interdot charge transition with $T_{\rm lat}$ = 10 mK. Dashed lines mark the boundaries of the DQD charge stability diagram and the black arrow denotes the detuning parameter $\epsilon$. (b) DQD energy level diagram with $E_L = E_R = 51$ $\mu$eV, $t = 20.9$ $\mu$eV and $t' = 15.0$ $\mu$eV. The grey dashed lines show the uncoupled energy levels ($t = t' = 0$). (c) $A/A_0$ measured as a function of $\epsilon$ for three values of $V_\text{MB}$, and fits to theory (black lines).}
	\label{fig:2}
\end{figure}

We now focus on the $(1,0) \leftrightarrow (0,1)$ interdot charge transition. Figure~\ref{fig:2}(a) shows $A/A_0$ as a function of $V_\text{LP}$ and $V_\text{RP}$, with $V_\text{MB} = 323$ mV. A clear reduction in $A/A_0$, to a minimum value of $\sim$0.8 (green arrows), is seen along the interdot charge transition where $\epsilon$ = 0. Parallel to this central minimum, two additional minima in $A/A_0$ are also visible (red and blue arrows). The observed cavity response is strikingly different from previously reported devices, where $A/A_0$ exhibited either a single minimum at $\epsilon = 0$ for $2t_{\rm c}/h > f_\text{c}$, or two minima with similar depths at values of $\epsilon$ where $\sqrt{\epsilon^2+4t_{\rm c}^2}/h = f_\text{c}$ when $2t_{\rm c}/h < f_\text{c}$ \cite{Walraff_2012_PRL,Petersson_Nature_2012,Graphene_cQED_2015,Kontos_PRB_2014,Mi_Science_2016}. Here $t_{\rm c}$ is the interdot tunnel coupling. These additional features suggest the presence of higher-lying avoided crossings in the DQD energy level diagram that lead to a non-zero electric susceptibility at finite values of $\epsilon$.

A qualitative understanding of the data can be obtained considering the full DQD energy diagram shown in Fig.~\ref{fig:2}(b) \cite{PhysRevB.94.195305}. Here the DQD is modeled as a four-level system consisting of the left dot ground state $\ket{L} = \ket{(1,0)}$, left dot excited state $\ket{L'} = \ket{(1',0)}$, right dot ground state $\ket{R} = \ket{(0,1)}$ and right dot excited state $\ket{R'} = \ket{(0,1')}$. For large detuning $|\epsilon| > 100$ $\mu$eV, valley states within the same dot are separated by the respective valley splitting, $E_L$ and $E_R$. For small detuning $|\epsilon| < 100$ $\mu$eV, the four states are hybridized by the intra-valley tunnel coupling $t$ and the inter-valley tunnel coupling $t'$, giving rise to a total of four avoided crossings \cite{PhysRevB.94.195305}. The strong minimum in $A/A_0$ at $\epsilon = 0$ is predominantly due to the avoided crossing involving the DQD ground states $\ket{L}$ and $\ket{R}$ (green arrows), similar to previous work \cite{Mi_APL_2016,Walraff_2012_PRL,Petersson_Nature_2012,Kontos_PRB_2014,Graphene_cQED_2015,Mi_Science_2016}. The two minima in $A/A_0$ at $\epsilon \neq 0$ are due to the avoided crossings involving states $\ket{L}$ - $\ket{R'}$ (red arrows) and $\ket{L'}$ - $\ket{R}$ (blue arrows), located at $\epsilon \approx \pm 50$ $\mu$eV. The lower visibility of these two minima arises from the smaller thermal population of the excited states. The $\ket{L'}$ - $\ket{R'}$ avoided crossing is expected to have no appreciable contribution to the cavity response due to the negligible population of the two highest-lying states. Moreover, for $E_L$ $\approx$ $E_R$, the $\ket{L'}$ - $\ket{R'}$ avoided crossing occurs near $\epsilon$ = 0 and its response would be masked by the $\ket{L}$ - $\ket{R}$ avoided crossing. The temperature dependence of the cavity response will be examined in more detail in Fig.\ 3.

A more quantitative data set is obtained by measuring $A/A_0$ as a function of $\epsilon$ for several values of $V_\text{MB}$, which primarily tunes $t$ and $t'$ [Fig.~\ref{fig:2}(c)]. With $V_\text{MB} = 327$ mV, the ``side minima'' have depths comparable to the central minimum at $\epsilon = 0$. When $V_\text{MB}$ is lowered to 323 mV, the central minimum becomes deeper, whereas the side minima remain relatively unchanged. As $V_\text{MB}$ is further lowered to 318 mV, the central minimum is split into two minima due to the fact that the $\ket{L}$ - $\ket{R}$ transition frequency now crosses the cavity frequency $f_\text{c}$ twice as $\epsilon$ is swept across the interdot charge transition \cite{Walraff_2012_PRL,Petersson_Nature_2012,Kontos_PRB_2014,Mi_Science_2016}.

The theory developed in Ref.\ \cite{PhysRevB.94.195305} is used to analyze these data. Starting from the four-level system shown in Fig.\ 2(b), the electric susceptibility $\chi$ is calculated as a function of $\epsilon$, and used to predict the cavity response $A/A_0$ with cavity input-output theory \cite{Petersson_Nature_2012}. We simultaneously fit the three data sets shown in Fig.\ 2(c) to theory, assuming that changes in $V_\text{MB}$ only modify $t$ and $t'$ \cite{PhysRevB.94.195305}. The fits to the data are plotted as black lines in Fig.~\ref{fig:2}(c) and are in excellent agreement with the experimental data with best fit valley splittings $E_\text{L} = E_\text{R} = 51$ $\mu$eV, which are comparable to values reported from devices made on similar wafers \cite{Dave_DQD_APL}. The equal valley splitting of the DQD is intentionally achieved through device tuning, and data showing $E_\text{L} \neq E_\text{R}$ are included in the Supplemental Material \cite{Supplement}. Best fit values of $t$ and $t'$ are listed in Fig.~\ref{fig:2}(c). Other inputs to the theory are the charge-cavity coupling rate $g_0 / 2 \pi = 19$ MHz, total charge decoherence rate $\gamma / 2 \pi = 30$ MHz, and the electron temperature $T_\text{e} = 135$ mK (at the cryostat base temperature $T_{\rm lat}$ = 10 mK). Low frequency charge noise is also accounted for in the model by smoothing $A/A_0$ using a Gaussian with standard deviation $\sigma_\epsilon = 1.5$ $\mu$eV \cite{Petersson_Nature_2012}. Fits to the cavity phase response (not shown) yield similar results. 

\begin{figure}[t]
	\centering
	\includegraphics[width=\columnwidth]{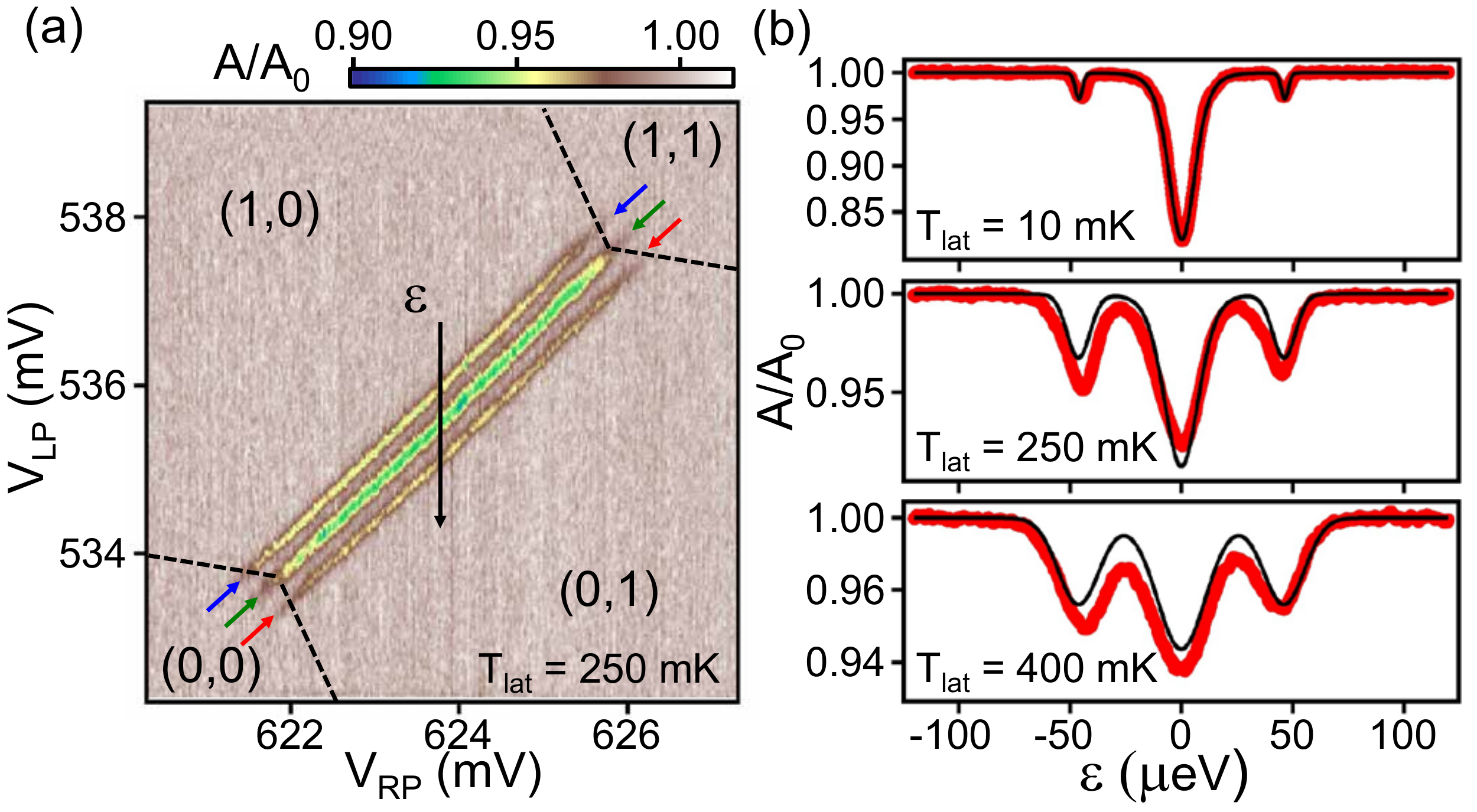}
	\caption{(Color Online). (a) Cavity transmission amplitude $A/A_0$ plotted as a function of $V_\text{LP}$ and $V_\text{RP}$ with $V_\text{MB} = 323$ mV and $T_{\rm lat}$ = 250 mK. (b) $A/A_0$ as a function of $\epsilon$ for three values of $T_{\rm lat}$, with $V_\text{MB}$ = 323 mV. Black solid lines are theory predictions. Data and fit for $T_{\rm lat}$ = 10 mK [Fig.~\ref{fig:2}(c)] are reproduced here for direct comparison.}
	\label{fig:3}
\end{figure}

Since only occupied electronic states will contribute to the electric susceptibility $\chi$, the visibility of the valley-induced features is likely limited by the relatively low electron temperature $k_{\rm B} T_\text{e} \ll E_{\rm L} \approx E_{\rm R}$, where $k_{\rm B}$ is Boltzmann's constant \cite{PhysRevB.94.195305}. This visibility may therefore be improved by raising the temperature of the DQD. Figure~\ref{fig:3}(a) shows the DQD stability diagram taken at $T_{\rm lat}$ = 250 mK. In comparison with Fig.~\ref{fig:2}(a), the side minima (red and blue arrows) are notably more visible relative to the central minimum. All three minima are also broader. These observations are expected since by raising temperature, the thermal population of the DQD excited states is increased and the ground state population is decreased. This leads to a smaller $\chi$ for the $\ket{L}$ - $\ket{R}$ avoided crossing, but a larger $\chi$ for the $\ket{L'}$ - $\ket{R}$ and $\ket{L}$ - $\ket{R'}$ avoided crossings \cite{PhysRevB.94.195305,Schroer_Parity_PRL}. As a result, the difference between the depths of the side minima and the central minimum in $A/A_0$ is reduced. On the other hand, background charge noise is also expected to increase at higher temperatures, leading to broadening of the observed features \cite{PhysRevLett.110.146804}.

Close agreement between theory and experiment is demonstrated by the plots in Fig.~\ref{fig:3}(b), where we show $A/A_0$ as a function of $\epsilon$ for $T_{\rm lat}$ = 10, 250, and 400 mK. The temperature dependence of the cavity response is theoretically modeled by taking into account the Fermi-Dirac distribution of the electrons in the leads and the Bose-Einstein distribution of the phonon bath, which are assumed to be in equilibrium with the base temperature of the cryostat $T_{\rm lat}$, except for $T_{\rm lat}$ = 10 mK where the best fit is found with an electron temperature $T_{\rm e}$ = 135 mK. The temperature-dependent charge noise used in the model is $\sigma_\epsilon = 1.5$ $\mu$eV for $T_{\rm lat} = 10$ mK, $\sigma_\epsilon = 5$ $\mu$eV for $T_{\rm lat} = 250$ mK and $\sigma_\epsilon = 9$ $\mu$eV for $T_{\rm lat} = 400$ mK~\cite{dial_charge_2013}. The charge decoherence rate $\gamma/2\pi$ has little impact on theoretical predictions in Fig.~\ref{fig:3}(b) and is fixed at 30 MHz. In comparison with the data at $T_{\rm lat}$ = 10 mK, we conclude that raising device temperature is an effective method for improving the relative visibility of valley-induced dispersive features.

\begin{figure}[t]
	\centering
	\includegraphics[width = \columnwidth]{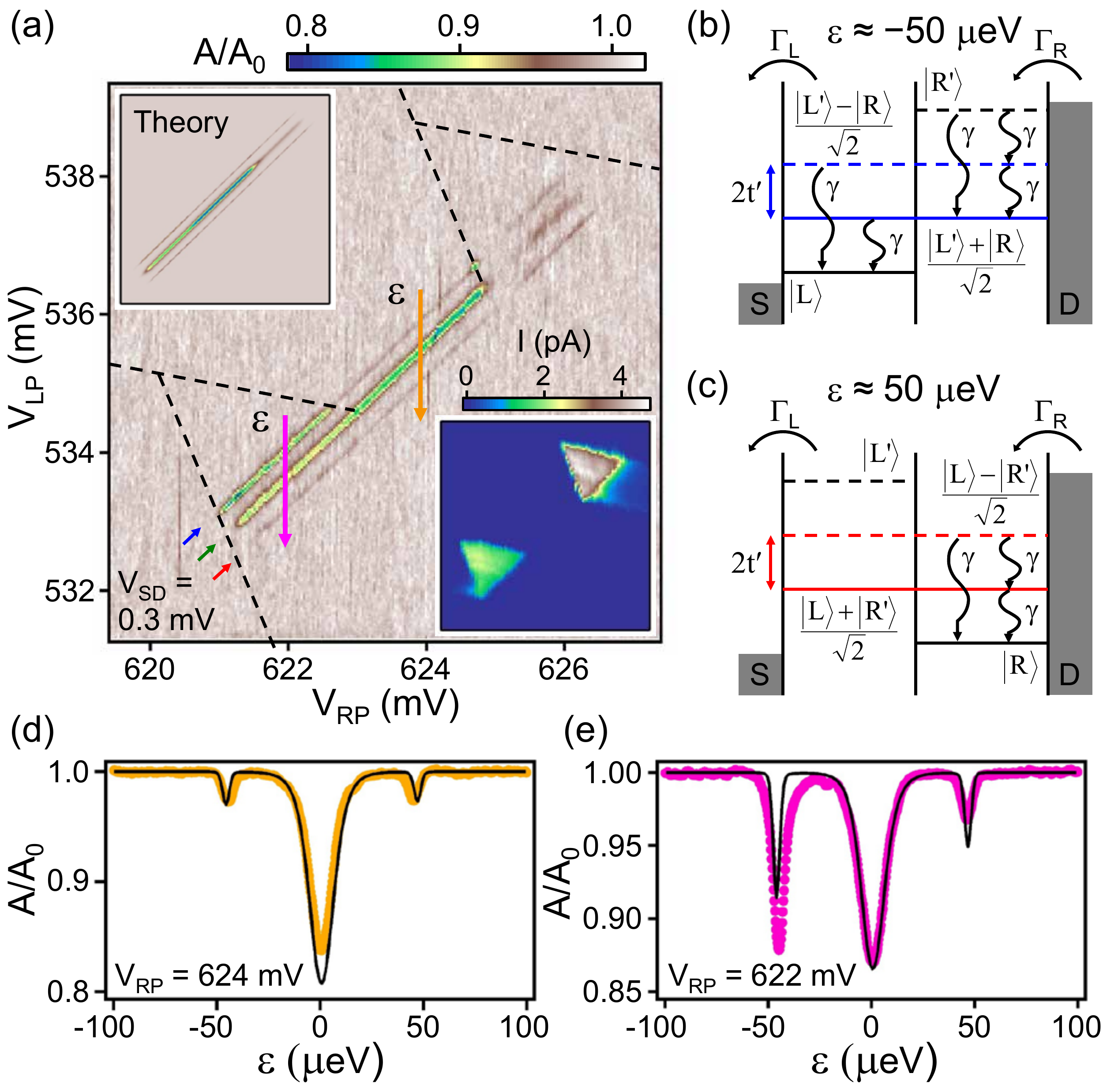}
	\caption{(Color Online). (a) $A/A_0$ measured with $V_\text{MB} = 323$ mV and $V_\text{SD} = 0.3$ mV. Upper inset: theory. Lower inset: Measured DQD current, $I$. Dashed lines mark the boundaries of the finite bias triangles. (b, c) DQD energy level diagrams at $V_\text{RP} = 622$ mV for (b) $\epsilon \approx -50$ $\mu$eV and (c) $\epsilon \approx 50$ $\mu$eV. (d, e) $A/A_0$ measured at (d) $V_\text{RP} = 624$ mV and (e) $V_\text{RP} = 622$ mV. Black solid lines are fits to theory.}
	\label{fig:4}
\end{figure}

While increasing temperature is effective at repopulating the higher-lying valley states, its effects are complicated by an overall increase in the level of charge noise in the device. We now show that a finite source-drain bias $V_{\rm SD}$ can effectively be used to repopulate the valley states and increase their visibility. The lower inset of Fig.\ \ref{fig:4}(a) shows the current through the DQD, $I$, as a function of $V_\text{LP}$ and $V_\text{RP}$ with $V_\text{SD} = 0.3$ mV. These data exhibit characteristic finite bias triangles (FBT), as expected for charge transport in a DQD \cite{Petta_RevMod}. The main panel of Fig.\ \ref{fig:4}(a) shows $A/A_0$ measured over the same range of gate voltages. $A/A_0$ is observed to vary appreciably along the base leg of the FBT, where $\epsilon \approx 0$. In the region between the FBTs, sequential single electron transport is forbidden due to Coulomb blockade. Here $A/A_0$ as a function of $\epsilon$ is little changed from the $V_\text{SD} = 0$ data [see Fig.~\ref{fig:2}(a)]. Within each FBT, electronic transport leads to non-equilibrium populations of the DQD electronic states and $A/A_0$ is strongly altered \cite{Kontos_PRB_2014}. To give a specific example, we observe an enhancement of the absolute visibility of the $\ket{L'}$ - $\ket{R}$ avoided crossing (blue arrow) in the lower FBT [Fig.~\ref{fig:4}(a)].

We can qualitatively understand the nonequilibrium cavity response by considering the transport process within the lower FBT. At $\epsilon \approx -50$ $\mu$eV [Fig.~\ref{fig:4}(b)], the relevant states involved are labeled in the figure, with $(\ket{L'} \pm \ket{R})/\sqrt{2}$ being hybridized valley-orbit states extended over the two dots. The charge transport cycle starts with the $\ket{0, 0}$ state, from which an electron tunnels from the right reservoir at a rate $\Gamma_R$ into one of the three states available in the right dot. Relaxation processes within the DQD (black curly arrows labeled with rate $\gamma$) lead to one of the three states available in the left dot, from which the electron may tunnel with a rate $\Gamma_L$ onto the left lead. This transport cycle results in an increased population difference between the hybridized valley-orbit states compared to equilibrium and thus a high visibility of the $\ket{L'}$ - $\ket{R}$ avoided crossing. In contrast, at the opposite detuning [$\epsilon \approx 50$ $\mu$eV, see Fig.~\ref{fig:4}(c)], an electron may relax into state $\ket{R}$ where it will remain stuck. Consequently, the population difference between the hybridized valley-orbit states $(\ket{L} \pm \ket{R'})/\sqrt{2}$ is not increased and no enhancement in the visibility of the $\ket{L}$ - $\ket{R'}$ avoided crossing is observed.

To quantitatively model the nonequilibrium cavity response, we solve for the steady state occupation probability $\rho_k$ of each DQD eigenstate $\ket{k}$ (see Supplemental Material for details \cite{Supplement}). Using a Lindblad master equation approach \cite{breuer_theory_2007}, we derive a set of rate equations for $\rho_k$: $0=\dot{\rho}_k=\sum_{j\neq k} \left( \Gamma_{kj} \rho_j - \Gamma_{jk} \rho_k \right) + \sum_{j\neq k} \left( \tau_{kj} \rho_j - \tau_{jk} \rho_k \right)$. Here $\Gamma_{jk}=\sum_{v = \pm,l=\{L,R\}} \Gamma_l \left(\left|\left\langle j | 
c_{l,v} | k \right\rangle \right|^2 + \left|\left\langle k | c_{l,v} | j \right\rangle \right|^2\right) n_{jk}^{(l)}$ denotes the rate at which the state $k$ transits to state $j$ due to an electron tunneling on or off the DQD, with $\Gamma_l$ being the tunneling rate to lead $l$, $c_{l,v}$ the annihilation operator for electrons in dot $l$ and valley $v$ and the $n_{jk}^{(l)}$ factors account for the finite temperature in the source-drain leads. The sum $\sum_{j\neq k} \left( \tau_{kj} \rho_j - \tau_{jk} \rho_k \right)$ describes decay processes between states with the same total number of electrons, where the decay rate $\tau_{jk} \sim \gamma$ for states with one electron in the DQD. The resulting values of $\rho_k$ are then used to calculate $A/A_0$ via cavity input-output theory \cite{PhysRevB.94.195305}.

As shown in Figs.~\ref{fig:4}(d, e), theory and the experimental data are in good agreement, with best fit tunneling rates $\Gamma_L = 62$ MHz, $\Gamma_R = 132$ MHz and $\gamma = 188$ MHz. In the upper inset of Fig.~\ref{fig:4}(a), we have calculated $A/A_0$ over the same gate voltage range as the data using these parameters and a capacitance matrix based model of a DQD charge stability diagram \cite{Supplement,Kouwenhoven_Review_2002}. The theoretical cavity response agrees well with the data. Lastly, enhancement in the absolute visibility of the $\ket{L}$ - $\ket{R'}$ avoided crossing (red arrows) is seen in the upper FBT with $V_\text{SD} = -0.3$ mV (data shown in the Supplemental Material \cite{Supplement}).

In conclusion, we observe dispersive features in the cavity response of a hybrid Si/SiGe DQD-cQED device that arise from the valley degree of freedom in Si. As predicted by theory, the cavity response is sensitive to the valley splitting in each dot, the inter- and intra-valley tunnel couplings, and the time-averaged occupation of the levels \cite{PhysRevB.94.195305}. The relative occupation of the DQD energy levels can be driven out of equilibrium by increasing temperature or applying a source-drain bias, thereby increasing the visibility of the valley states. These measurements constitute an efficient method for accurate spectroscopy of valley states in Si/SiGe heterostructures and may be readily applied to other Si devices, such as Si metal-oxide-semiconductor (MOS) quantum dots. Rapid and accurate measurements of the valley splitting may accelerate progress toward understanding, and ultimately controlling valley splittings in these material systems which are highly relevant to spin-based quantum information processing.

\begin{acknowledgements}
Supported by Army Research Office grant W911NF-15-1-0149 and the Gordon and Betty Moore Foundation’s EPiQS Initiative through grant GBMF4535. This material is based on work supported by the U.S. Department of Defense under contract H98230-15-C0453. Devices were fabricated in the Princeton University Quantum Device Nanofabrication Laboratory.
\end{acknowledgements}

\bibliographystyle{apsrev4-1}
\bibliography{references_v6}

\end{document}